\documentclass{elsarticle}
\usepackage{graphicx}
\input{tcilatex}
\begin{document}

\begin{frontmatter}
\title{Solution to Banff 2 Challenge Based on Likelihood Ratio Test}
\author[wolf]{Wolfgang A. Rolke}
\address[wolf]{Department of Mathematics, University of Puerto Rico - Mayag\"{u}ez, Mayag\"{u}ez, PR 00681, USA, 
\newline Postal Address: PO Box 3486, Mayag\"{u}ez, PR 00681, 
\newline Tel: (787) 255-1793, Email: wolfgang@puerto-rico.net}

\begin{abstract}
   We describe our solution to the Banff 2 challenge problems as well as the outcomes.
\end{abstract}

\end{frontmatter}\newpage

\section{Introduction}

In July of 2010 a conference was held on the statistical issues relevant to
significance of discovery claims at the LHC. The conference took place at
the Banff International Research Station in Banff, Alberta, Canada. After
many discussions it was decided to hold a competition to see which methods
would perform best. One of the participants, Thomas Junk, would create a
large number of data sets, some with a signal and some without. There were
two main parts to the competition:

Problem 1 was essentially designed to see whether the methods could cope
with the "look-elsewhere" effect, the issue of searching through a mass
spectrum for a possible signal.

Problem 2 was concerned with the problem that sometimes there are no known
distributions for either the backgrounds or the signal and they have to be
estimated via Monte Carlo.

For a detailed description of the problems as well as the data sets and a
discussion of the results see Tom Junk's CDF web page at
http://www-cdf.fnal.gov/\symbol{126}trj/. In this paper we will present a
solution based on the likelihood ratio test, and discuss the performance of
this method in the challenge.

\section{The method}

Our solution for both problems is based on the likelihood ratio test
statistic

\[
\lambda (\mathbf{x})=2\left( \max \{\log L(\theta |\mathbf{x):\theta \}-}%
\max \{\log L(\theta |\mathbf{x):\theta \in \Theta }^{0}\mathbf{\}}\right) 
\]

According to standard theorems in Statistics $\lambda (\mathbf{X})$ often
has a $\chi ^{2}$ distribution with the number of degrees of freedom the
difference between the number of free parameters and the number of free
parameters under the null hypothesis. This turns out to be true for problem
2 but not for problem 1, in which case the null distribution can be found
via simulation.

\subsection{Problem 1}

Here we have:%
\[
\begin{tabular}{l}
$f(x)=10.00045e^{-10x}$, \ $0<x<1$ \\ 
$\varphi (x;E)=\frac{1}{\sqrt{2\pi }0.03}e^{-\frac{1}{2}\frac{(x-E)^{2}}{%
0.03^{2}}}$ \\ 
$g(x;E)=\frac{\varphi (x;E)}{\int_{0}^{1}\varphi (t;E)dt}$, \ $0<x<1$ \\ 
$h(x;\alpha ,E)=(1-\alpha )f(x)+\alpha g(x;E)$ \\ 
$H_{0}:\alpha =0$ vs $H_{a}:\alpha >0$ \\ 
$\log L(\alpha ,E|\mathbf{x)=}\sum_{i=1}^{n}\log \left[ (1-\alpha
)f(x_{i})+\alpha g(x_{i};E)\right] $%
\end{tabular}%
\]

Now $\max \{\log L(\alpha ,E|\mathbf{x)\}}$ is the log likelihood function
evaluated at the maximum likelihood estimator and $\max \{\log L(\alpha ,E|%
\mathbf{x):\mathbf{\theta \in \Theta }^{0}\}=}\log L(0,0|\mathbf{x)}$. Note
that if $\alpha =0$ any choice of E yields the same value of the likelihood
function.

In the following figure we have the histogram of 100000 values of $\lambda (%
\mathbf{x})$ for a simulation with $n=500$ and $\alpha =0$ together with the
densities of the $\chi ^{2}$ distribution with df's from 1 to 5. Clearly non
of these yields an acceptable fit. Instead we use the simulated data to find
the 99\% quantile and reject the null hypothesis if $\lambda (\mathbf{x})$
is larger than that, shown as the vertical line in the graph.

\begin{figure}
	\centering
		\includegraphics[width=0.90\textwidth]{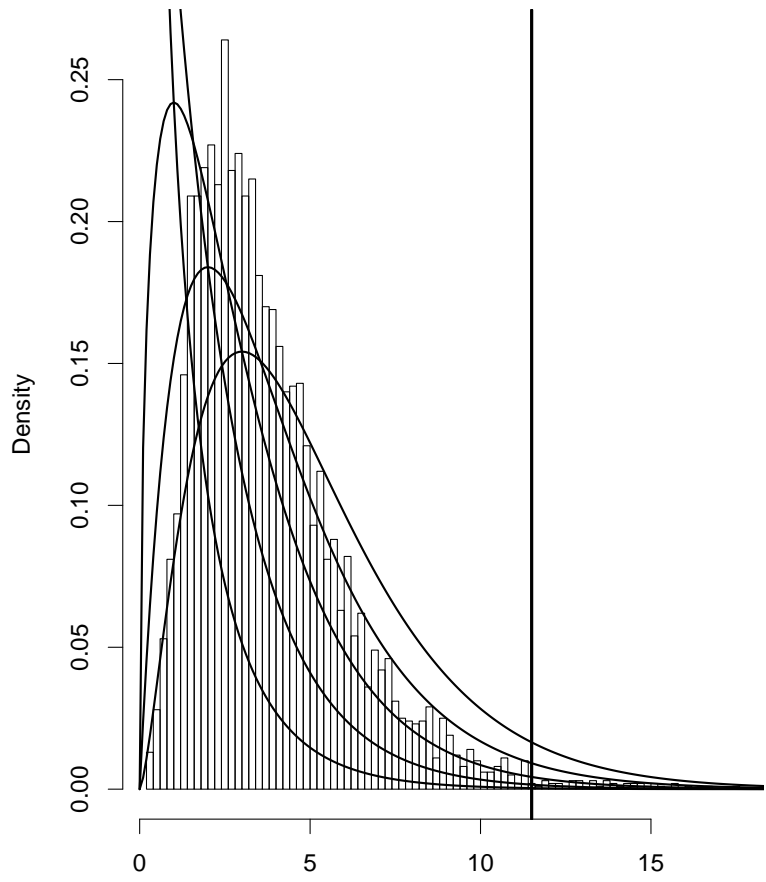}
	\caption{Histogram of $100000$ values of the null distribution, with several fits from chi-square distributions and $99^{th}$ percentile.}
	\label{fig:fig1}
\end{figure}

In general the critical value will depend on the sample size, but for those
in the challenge ($500-1500$) it is always about $11.5$.

If it was decided to do discovery using $5\sigma $ the critical value can be
found using importance sampling. Recently Eilam Gross and Ofer Vitells have
developed an analytic upper bound for the tail probabilities of the null
distribution, see "\textit{Trial factors for the look elsewhere effect in
high energy physics", }Eilam Gross, Ofer Vitells,
Eur.Phys.J.C70:525-530,2010. Their result agrees with our simulations.

Finding the mle is a non-trivial exercise because there are many local
minima. The next figure shows the log-likelihood as a function of $E$ with $%
\alpha $ fixed at $0.05$ for 4 cases.

\begin{figure}
	\centering
		\includegraphics[width=0.90\textwidth]{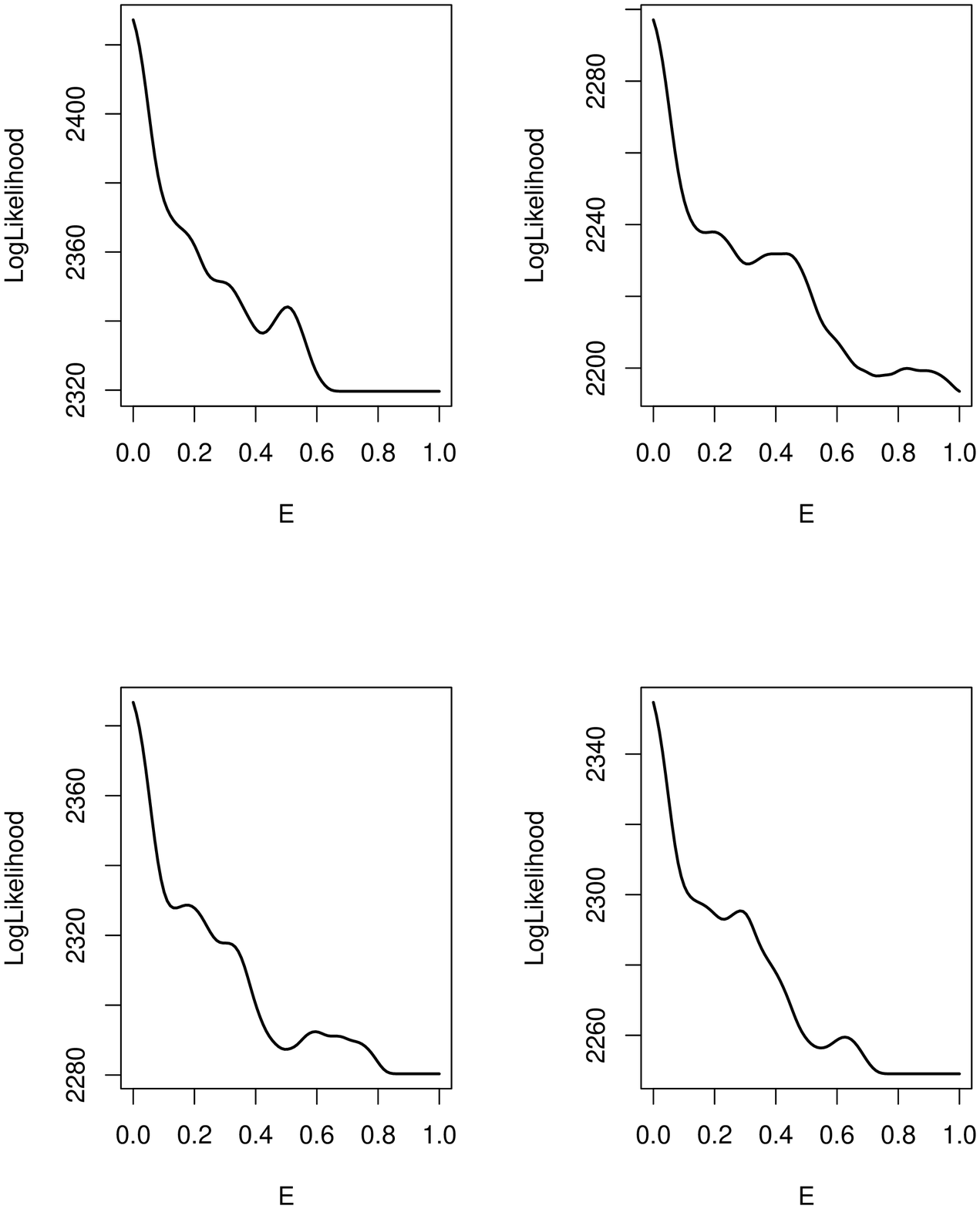}
	\caption{Log-Likelihood as a function of signal location $E$. $\alpha$=0.05}
	\label{fig:fig2}
\end{figure}

To find the mle we used a two-step procedure: first a fine grid search over
values of $E$ from $-0.015$ to $1$ in steps of $0.005$. At each value of $E$
the corresponding value of $\alpha $ that maximizes the log-likelihood is
found. In a second step the procedure starts at the best point found above
and uses Newton-Raphson to find the overall mle.

\subsection{Problem 2}

Again we want to use:%
\[
\begin{tabular}{l}
$h(x;\alpha ,\beta )=(1-\alpha -\beta )f_{1}(x)+\beta f_{2}(x)+\alpha g(x)$
\\ 
$H_{0}:\alpha =0$ vs $H_{a}:\alpha >0$ \\ 
$\log L(\alpha ,\beta |\mathbf{x)=}\sum_{i=1}^{n}\log \left[ (1-\alpha
-\beta )f_{1}(x)+\beta f_{2}(x)+\alpha g(x)\right] $%
\end{tabular}%
\]

Now $\max \{\log L(\alpha ,\beta |\mathbf{x)\}}$ is the log likelihood
function evaluated at the maximum likelihood estimator and $\max \{\log
L(\alpha ,\beta |\mathbf{x):\mathbf{\theta \in \Theta }^{0}\}}=\max \mathbf{%
\{}\log L(0,\beta |\mathbf{x)}:\beta \}$.

The difficulty is of course that we don't know $f_{1}$, $f_{2}$ or $g$. We
have used three different ways to find them:

\subsubsection{Parametric Fitting}

Here one tries to find a parametric density that gives a reasonable fit to
the data. For the data in the challenge this turns out to be very easy. In
all three cases a Beta density gives a very good fit:

\begin{figure}
	\centering
		\includegraphics[width=0.90\textwidth]{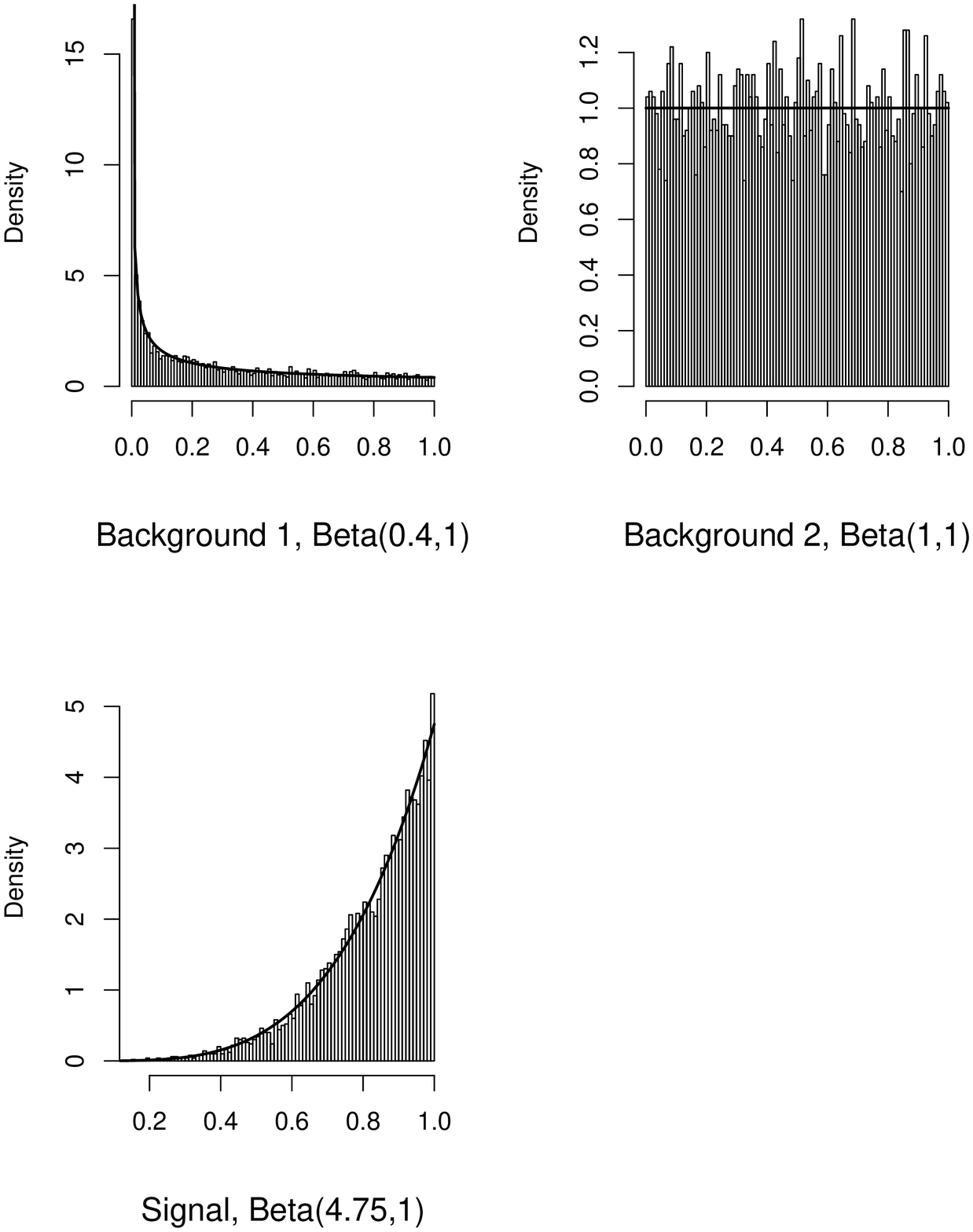}
	\caption{Parametric fits to the three MC samples.}
	\label{fig:fig3}
\end{figure}

\subsubsection{Nonparametric Fitting:}

There are a variety of methods known in Statistics for non-parametric
density estimation. The difficulty with the data in the challenge is that it
is bounded on a finite interval, a very common feature in HEP data. Moreover
the slope of the density of Background 1 at 0 is infinite. I checked a
number of methods and eventually ended up using the following: for
Background 2, the Signal and the right half of Background 1 i bin the data
(250 bins) find the counts and scale them to integrate to unity. Then i use
the non-parametric density estimator loess from R with the default span
(smoothing parameter). This works well except on the left side of Background
1. There the infinite slope of the density would require a smoothing
parameter that goes to 0. Instead i transform the data with $\log \frac{x}{%
1-x}$. The resulting data has a density without boundary, which i estimate
using the routine density from R, again with the default bandwidth. This is
then back-transfomed to the 0-1 scale. This works well for the left side but
not the right one, and so i "splice" the two densities together in the
middle. The resulting densities are shown here:

\begin{figure}
	\centering
		\includegraphics[width=0.90\textwidth]{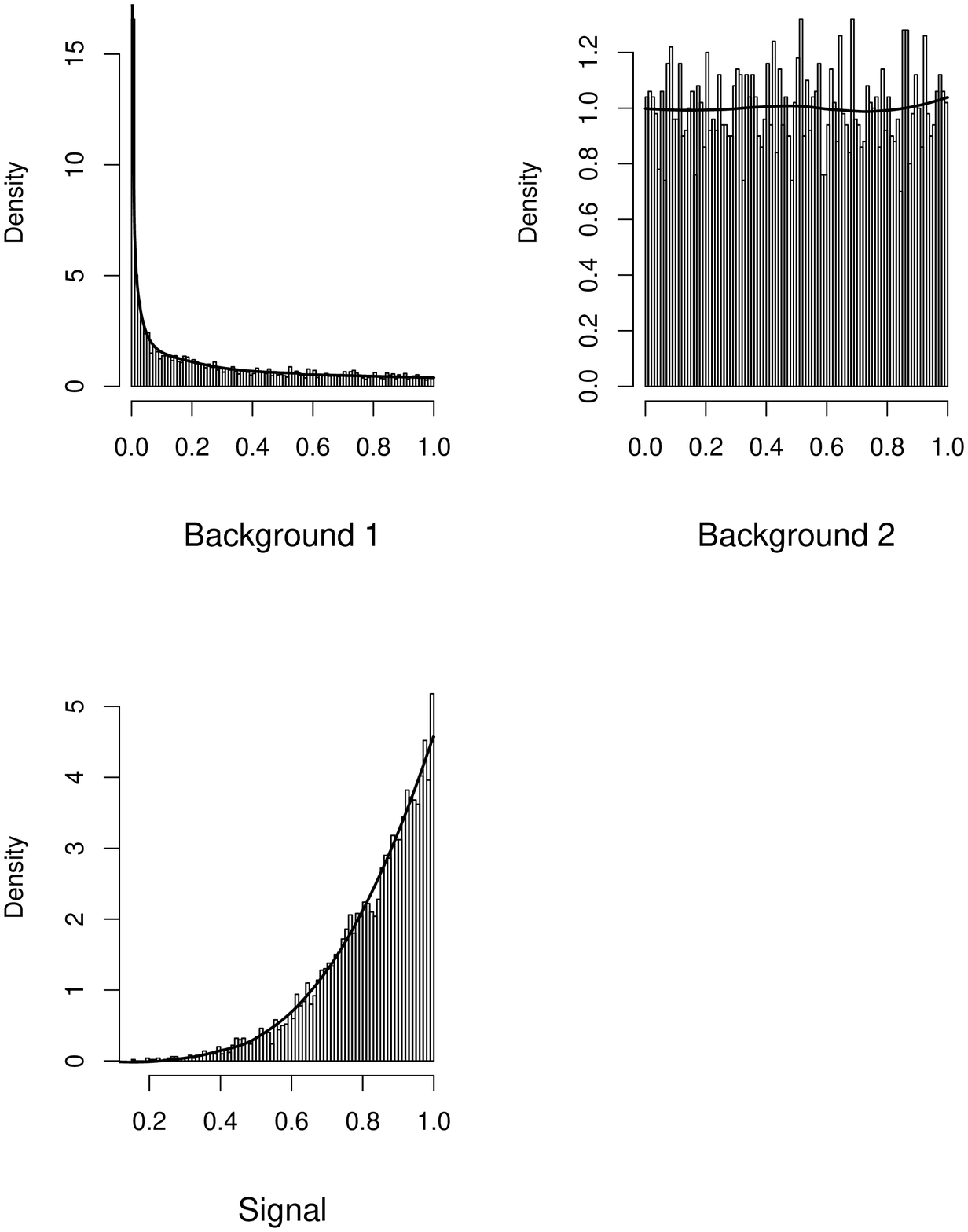}
	\caption{Non-parametric fits to the three MC samples.}
	\label{fig:fig4}
\end{figure}

\subsubsection{Semiparametric Fitting}

It is possible to combine the two approaches above: fit some of the data
parametrically and others non-parametrically, for example if the signal is
known to have a Gaussian distribution but the background density is Monte
Carlo data.

\bigskip

For the data in the challenge the three methods give very similar results,
so i am submitting only the solution using the parametric fits.

\subsubsection{Back to the Test}

What is the null distribution of $\lambda (\mathbf{x})$ now? In the
following figure we have the histogram of 5000 values of $\lambda (\mathbf{x}%
)$ for a simulation with $500$ events from Background 1, $100$ events from
Background 2 and no Signal events. $\alpha =0$ together with the density of
a $\chi ^{2}$ distribution with $1$ df. The densities are fit parametrically.
\begin{figure}
	\centering
		\includegraphics[width=0.90\textwidth]{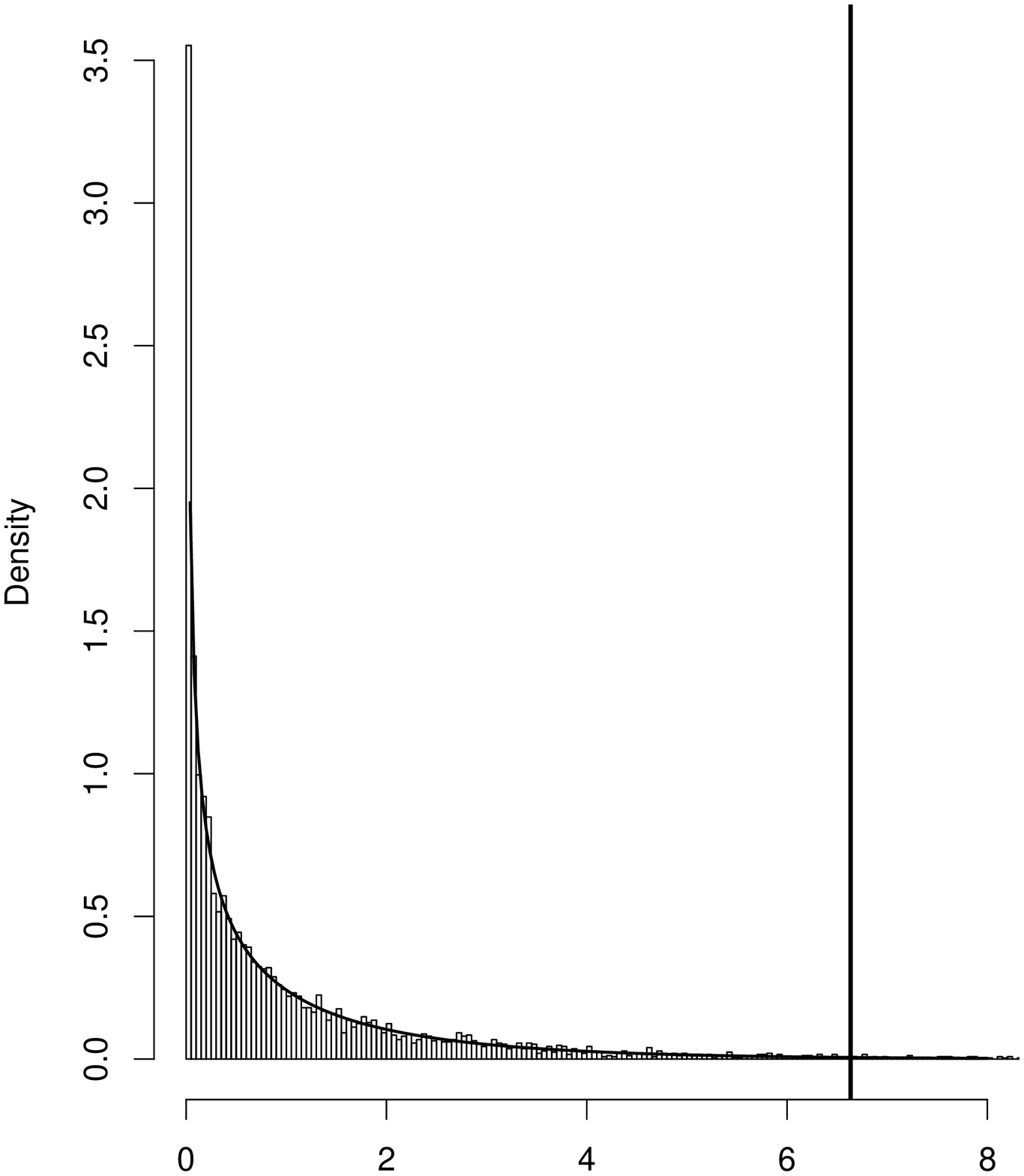}
	\caption{Histogram of null distribution for problem 2, with chi-square density.}
	\label{fig:fig5}
\end{figure}

Clearly this yield a very good fit, so we will reject the null hypothesis if 
$\lambda (\mathbf{x})>q\chi ^{2}(0.99,1)=6.635$, the $99^{th}$ percentile of
a chi-square distribution with $1$ degree of freedom. The same result holds
if the fitting was done non parametrically or semi parametrically.

\subsection{Error Estimation}

We have the following large sample theorem for maximum likelihood
estimators: if we have a sample $X_{1},..,X_{n}$ with density $f(x;p)$ and $%
\widehat{p}$ is the mle for the parameter $p$, then under some regularity
conditions%
\[
\sqrt{n}(\widehat{p}-p)\symbol{126}N(0,\sigma ) 
\]%
where $\sigma ^{2}=-1/nE[\frac{d^{2}}{dp^{2}}\log f(X;p)]$, the Fisher
information number. This can be estimated using the observed Fisher
information number $\frac{1}{n}\sum_{i=1}^{n}\frac{d^{2}}{dp^{2}}\log
h(X_{i};p).$

For problem 1 we find:

\[
\begin{tabular}{l}
Signal density: $S(x;E)=\frac{g(x;E)}{q(E)}$ \\ 
$g(x;E)=dn(x;E,\sigma )=\frac{1}{\sqrt{2\pi }\sigma }e^{-\frac{1}{2}\frac{%
(x-E)^{2}}{\sigma ^{2}}}$ \\ 
$pn(x;E,\sigma )=\int_{-\infty }^{x}dn(t;E,\sigma )dt$ \\ 
$\varphi (x)=\frac{1}{\sqrt{2\pi }}e^{-\frac{1}{2}x^{2}}$ \\ 
$q(E)=1-pn(0;E,\sigma )=1-pn(-E/\sigma ;0,0)$ \\ 
$g_{E}=\frac{dg}{dE}=g\frac{x-E}{\sigma ^{2}}$ \\ 
$g_{EE}=\frac{d^{2}g}{dE^{2}}=g\left( \frac{x-E}{\sigma ^{2}}\right) ^{2}+g%
\frac{-1}{\sigma ^{2}}=g\left[ \left( \frac{x-E}{\sigma ^{2}}\right) ^{2}-%
\frac{1}{\sigma ^{2}}\right] $ \\ 
$\frac{d}{dE}q(E)=-\varphi (-E/\sigma )\left( -\frac{1}{\sigma }\right)
=\varphi (E/\sigma )/\sigma $ \\ 
$\frac{d}{dx}\varphi (x)=-x\varphi (x)$ \\ 
$r(E)=\log S(x;E)=\log g-\log q=const-\frac{(x-E)^{2}}{2\sigma ^{2}}-\log q$
\\ 
$r^{\prime }=\frac{x-E}{\sigma ^{2}}-\frac{\varphi (E/\sigma )/\sigma }{q}$
\\ 
$r^{\prime \prime }=-\frac{1}{\sigma ^{2}}+\frac{-\varphi (E/\sigma )\frac{E%
}{\sigma ^{2}}q-\varphi (E/\sigma )^{2}/\sigma ^{2}}{q^{2}}=$ \\ 
$-\frac{1}{\sigma ^{2}}\left[ 1+\varphi (E/\sigma )\frac{qE+\varphi
(E/\sigma )}{q^{2}}\right] $ \\ 
$\frac{dS}{dE}=Sr^{\prime }=S\left[ \frac{x-E}{\sigma ^{2}}-\frac{\varphi
(E/\sigma )/\sigma }{q}\right] $ \\ 
$\frac{d^{2}S}{dE^{2}}=S(r^{\prime \prime }+r^{\prime 2})=$ \\ 
$S\left[ -\frac{1}{\sigma ^{2}}\left[ 1+\varphi (E/\sigma )\frac{E\left[
1-\Phi (0,E)\right] +\varphi (E/\sigma )}{\left[ 1-\Phi (0,E)\right] ^{2}}%
\right] +\left( \frac{x-E}{\sigma ^{2}}+\frac{\varphi (E/\sigma )/\sigma }{%
1-\Phi (0,E)}\right) ^{2}\right] $ \\ 
$\psi (x;\alpha ,E)=(1-\alpha )f(x)+\alpha S(x;E)$ \\ 
$\sum_{i=1}^{n}\log \psi (x_{i};\alpha ,E)=$ \\ 
$\sum_{i=1}^{n}\log \left[ (1-\alpha )f(x_{i})+\alpha S(x_{i};E)\right] $ \\ 
$\frac{d\psi }{d\alpha }=S-f$ \\ 
$\frac{d\psi }{dE}=\alpha S^{\prime }=\alpha S\left[ \frac{x-E}{\sigma ^{2}}-%
\frac{\varphi (E/\sigma )/\sigma }{q}\right] $ \\ 
$\frac{d\log \psi }{d\alpha }=\frac{\psi _{\alpha }}{\psi }=\frac{S-f}{\psi }
$ \\ 
$\frac{d\log \psi }{dE}=\frac{\psi _{E}}{\psi }=\alpha \frac{S}{\psi }\left[ 
\frac{x-E}{\sigma ^{2}}-\frac{\varphi (E/\sigma )/\sigma }{q}\right] $ \\ 
$\frac{d^{2}\log \psi }{d\alpha ^{2}}=-\frac{\left( S-f\right) ^{2}}{\psi
^{2}}$ \\ 
$\frac{d^{2}\log \psi }{d\alpha dE}=\frac{S^{\prime }\psi -(S-f)\psi _{E}}{%
\psi ^{2}}$ \\ 
$\frac{d^{2}\log \psi }{dE^{2}}=\frac{d}{dE}\left[ \frac{\alpha S^{\prime }}{%
\psi }\right] =\frac{\alpha S^{\prime \prime }\psi -\psi _{g}^{2}}{\psi ^{2}}
$%
\end{tabular}%
\]

so the standard error of $\alpha $ is estimated with $\left[
\sum_{i=1}^{n}\left( \frac{S(X_{i};\widehat{\alpha },\widehat{E})-f(X_{i})}{%
\psi (X_{i};\widehat{\alpha },\widehat{E})}\right) ^{2}\right] ^{-1/2}$and
the standard error of $E$ is given by $\left[ \sum_{i=1}^{n}\left( \frac{%
\alpha S^{\prime \prime }\psi -\psi _{g}^{2}}{\psi ^{2}}\right) ^{2}\right]
^{-1/2}$.

For problem 2 the errors are found similarly.

How good are these errors? As always with large sample theory, there is a
question whether it works for a specific problem at the available sample
sizes. Here is the result of a mini MC study: we generate $500$ events from
the background of problem 1 and $4$ events from the signal with $E=0.8$.
This is repeated $2000$ times. The histogram of estimates for the mixing
ratio $\alpha $ is shown here:

\begin{figure}
	\centering
		\includegraphics[width=0.90\textwidth]{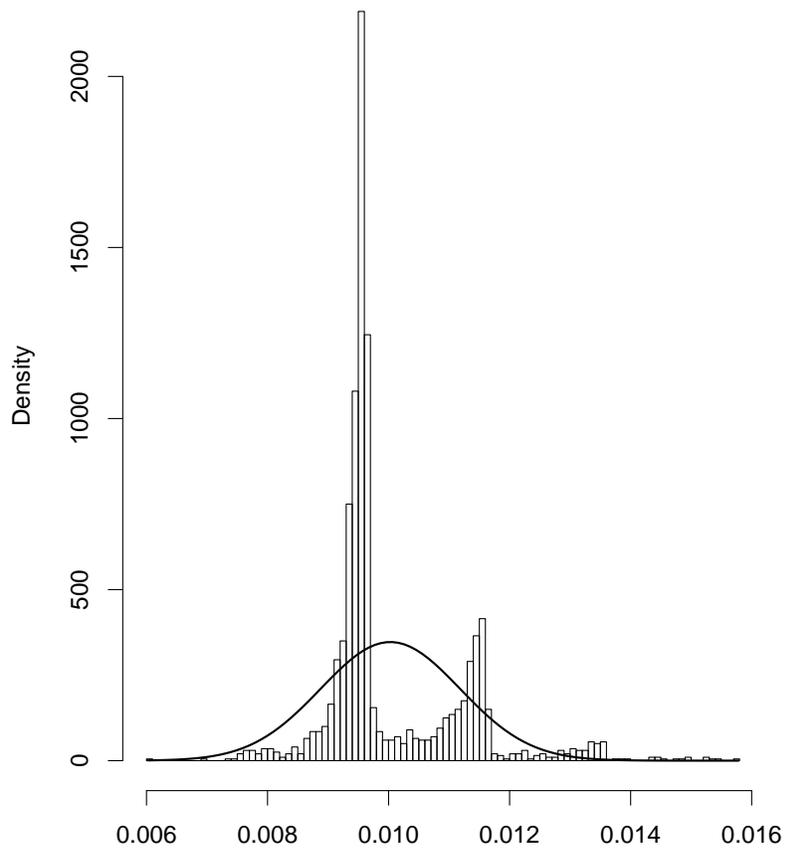}
	\caption{Histogram of estimates for the mixing ratio a. MC generated 500 events from the background of problem 1 and 4 events from the signal with E=0.8. This is repeated 2000 times.}
	\label{fig:fig6}
\end{figure}

Clearly the distribution of the estimates is not Gaussian, and therefore the
errors are wrong.

So we need a different method for finding the 68\% intervals. This can be
done via the statistical bootstrap. In our submission we will include
intervals based on the bootstrap, specifically the $16^{th}$ and the $84^{th}
$ percentile of a bootstrap sample of size $300$. In a real live problem it
would be easy to run a mini MC for a specific case and then decide which
errors to use. Because we have to process $20000$ data sets we can not do so
for each individually, and therefore use of the bootstrap intervals.

\subsection{Power Studies}

\paragraph{Problem 1:}

We used the following code to do the power study:

counter=0;

for(int k=0;k\TEXTsymbol{<}10000;++k) \{

\qquad nsig=rpois(0.07519885*D,seed);

\qquad for (int i=0; i\TEXTsymbol{<}nsig; ++i) x[i]=rnorm(E,seed);

\qquad nback=rpois(1000,seed);

\qquad for (int i=0; i\TEXTsymbol{<}nback; ++i) x[nsig+i]=rbackground(seed)

\qquad lrt=findLRT(x);

\qquad if(lrt\TEXTsymbol{>}11.5) ++counter;

\}

power=counter/M;

\bigskip

The results are :

$%
\begin{array}{ll}
(D,E) & \text{Power} \\ 
(1010,0.1) & \mathbf{0.356} \\ 
(137,0;0.5) & \mathbf{0.457} \\ 
(18,0;0,9) & \mathbf{0.184}%
\end{array}%
$

\paragraph{Problem 2:}

We used the following code to do the power study:

counter=0;

for(int k=0;k\TEXTsymbol{<}10000;++k) \{

\qquad n1=rnorm(1,900,90)

\qquad x1=sample(bc2p2b1mc,size=n1)

\qquad n2=max(rnorm(1,100,100),0)

\qquad x2=sample(bc2p2b2mc,size=n2)

\qquad n3=rpois(75,seed);

\qquad x3=sample(bc2p2sigmc,size=n3)

\qquad nback=rpois(1000,seed);

\qquad x=c(x1,x2,x3)

\qquad lrt=findLRT(x);

\qquad if(lrt\TEXTsymbol{>}6.635) ++counter;

\}

power=counter/M;

The result is a power of $\mathbf{88\%}$.

\section{Discussion of Results}

For a detailed discussion of the performance of all the submitted methods
see Tom Junk's CDF web page at http://www-cdf.fnal.gov/\symbol{126}trj .
Here we will discuss only the results of our method.

\subsection{Problem 1}

True type I error probability $1.03\%$

Missed signals: $53.7\%$

So the method achieves the desired type I error probability of $1\%$.

How about the errors? For the cases were a signal was claimed correctly the
nominal $68\%$ CI included the true number of signal events $59\%$ of the
time and the true signal location $63\%$, somewhat lower than desired. This
is unexpected because these errors were tested via simulation. For example,
for one of the cases in the data set, $E=0.38$ and $40$ signal events, the
true error rates are $86.7\%$ for the number of signal events and $67.5\%$
for the signal location. It turns out, though, that the error estimates are
quite bad in cases were the signal rate is very low. For example for the
case $E=0.38$ and $20$ signal events the true error rates are $53.6\%$ for
the number of signal events and $40.2\%$ for the signal location.

We also checked the errors based on the Fisher Information as described
above. Their performance was comparable to the bootstrap errors. The
conclusion is that for cases were the signal is small error estimates are
difficult.

\subsection{Problem 2}

True type I error probability $2.56\%$

Missed signals: $22.5\%$

\bigskip

So for this problem the type I error probability is too large. The reason
turns out to be the parametric fit. We used Beta densities for all
components, specifically $Beta(0.4,1)$ for background 1, $%
Beta(1,1)(=Unif(0,1))$ for background 2 and $Beta(4.75,1)$ for the signal.
These give excellent fits to the MC data provided. For example, generating $%
5000$ random variates from a $Beta(0.4,1)$, running the Kolmogorov-Smirnoff
test and repeating this procedure many times rejects the null hypothesis of
equal distributions only about $7\%$ of the time, at a nominal $5\%$ rate.
This problem was caused by the size of the MC samples. If instead of $5000$
MC events there had been $50000$ the same procedure as above would have
rejected the null hypothesis of equal distributions $98\%$ of the time, and
a better fitting density would have to be found. The real conclusion here is
that a very good density estimate is required to make this test work.

\end{document}